\documentclass[aoas,MSNbibl,nameyear,dvips]{arximspdf}
\usepackage{algorithm}
\usepackage{algorithmicx}
\usepackage{algpseudocode}
\usepackage{graphicx}
\usepackage{url,breakurl}

\doi{10.1214/14-AOAS755} %kopijuoti is PTS
\volume{8}
\issue{3}
\pubyear{2014}
\firstpage{1750}
\lastpage{1781}
\docsubty{FLA}

\makeatletter
\floatname{algorithm}{Procedure}
\def\by{\mathbf{y}}
\newcommand{\0}{\mathbf{0}}
\newcommand{\errorrv}{\mathcal{E}}
\def\bmu{\bolds{\mu}}
\def\bp{\mathbf{p}}
\def\bI{\mathbf{I}}
\newcommand{\pbar}{\bar{p}}
\newcommand{\bbeta}{\bolds{\beta}}
\def\bX{\mathbf{X}}
\makeatother

\begin{document}
\begin{frontmatter}

\title{Variable selection for BART: An application to~gene~regulation}
\runtitle{Variable selection for BART}

\begin{aug}
\author[A]{\fnms{Justin}~\snm{Bleich}\corref{}\ead[label=e1]{jbleich@wharton.upenn.edu}},
\author[A]{\fnms{Adam}~\snm{Kapelner}\ead[label=e2]{kapelner@wharton.upenn.edu}\thanksref{T1}},
\author[A]{\fnms{Edward~I.}~\snm{George}\ead[label=e3]{edgeorge@wharton.upenn.edu}} \and
\author[A]{\fnms{Shane~T.}~\snm{Jensen}\ead[label=e4]{stjensen@wharton.upenn.edu}}
\runauthor{Bleich, Kapelner, George and Jensen}
\affiliation{University of Pennsylvania}
\address[A]{Department of Statistics\\
The Wharton School\\
University of Pennsylvania\\
3730 Walnut Street, 400 JMHH\\
Philadelphia, Pennsylvania 19104-6304\\
USA\\
\printead{e1}\\
\phantom{E-mail: }\printead*{e2}\\
\phantom{E-mail: }\printead*{e3}\\
\phantom{E-mail: }\printead*{e4}}
\end{aug}
\thankstext{T1}{Supported by the National Science Foundation graduate research fellowship.}

% HISTORY:
\received{\smonth{10} \syear{2013}}
\revised{\smonth{3} \syear{2014}}

% ABSTRACT
%
\begin{abstract}
We consider the task of discovering gene regulatory networks, which are
defined as sets of genes and the corresponding transcription factors
which regulate their expression levels. This can be viewed as a
variable selection problem, potentially with high dimensionality.
Variable selection is especially challenging in high-dimensional
settings, where it is difficult to detect subtle individual effects and
interactions between predictors. Bayesian Additive Regression Trees
[\texttt{BART},
\textit{Ann. Appl. Stat.} \textbf{4} (2010) 266--298]
provides a novel nonparametric
alternative to parametric regression approaches, such as the lasso or
stepwise regression, especially when the number of relevant predictors
is sparse relative to the total number of available predictors and the
\mbox{fundamental} relationships are nonlinear. We develop a principled
permutation-based inferential approach for determining when the effect
of a selected predictor is likely to be real. Going further, we adapt
the \texttt{BART} procedure to incorporate informed prior information
about variable importance. We present simulations demonstrating that
our method compares favorably to existing parametric and nonparametric
procedures in a variety of data settings. To demonstrate the potential
of our approach in a biological context, we apply it to the task of
inferring the gene regulatory network in yeast (\emph{Saccharomyces cerevisiae}). We find that our \texttt{BART}-based procedure is best
able to recover the subset of covariates with the largest signal
compared to other variable selection methods. The methods developed in
this work are readily available in the \texttt{R} package \texttt
{bartMachine}.
\end{abstract}

% KEYWORDS
\end{frontmatter}

\setcounter{footnote}{1}
%s1 #&#
\section{Introduction}

An important statistical problem in many application areas is variable
selection: identifying the subset of covariates that exert influence on
a response variable. We consider the general framework where we have a
continuous response variable $\by$ and a large set of predictor
variables $\mathbf{x}_1,\ldots,\mathbf{x}_K$. We focus on variable
selection in the sparse setting: only a relatively small subset of
those predictor variables truly influences the response variable.

One such example of a sparse setting is the motivating application for
this paper: inferring the gene regulatory network in budding yeast
(\emph{Saccharomyces cerevisiae}). In this application, we have a
collection of approximately 40 transcription factor proteins (TFs) that
act to regulate cellular processes in yeast by promoting or repressing
transcription of specific genes. It is unknown which of the genes in
our yeast data are regulated by each of the transcription factors.
Therefore, the goal of the analysis is to discover the corresponding
network of gene--TF relationships, which is known as a \textit{gene
regulatory network}. Each gene, however, is regulated by only a small
subset of the TFs which makes this application a sparse setting for
variable selection. The available data consist of gene expression
measures for approximately 6000 genes in yeast across several hundred
experiments, as well as expression measures for each of the
approximately 40 transcription factors in those experiments [\citet{JenCheSto07}].

This gene regulatory network was previously studied in \citet
{JenCheSto07} with a focus on modeling the relationship between genes
and transcription factors. The authors considered a Bayesian linear
hierarchical model with first-order interactions. In high-dimensional
data sets, specifying even first-order pairwise interactions can
substantially increase the complexity of the model. Additionally, given
the elaborate nature of biological processes, there may be interest in
exploring nonlinear relationships as well as higher-order interaction
terms. In such cases, it may not be possible for the researcher to
specify these terms in a linear model {a~priori}. Indeed,
\citet{JenCheSto07} acknowledge the potential utility of such
additions, but highlight the practical difficulties associated with the
size of the resulting parameter space. Thus, we propose a variable
selection procedure that relies on the nonparametric Bayesian model,
Bayesian Additive Regression Trees [\texttt{BART}, \citet{Chipman10}]. \texttt{BART} dynamically estimates a model from the
data, thereby allowing the researcher to potentially identify genetic
regulatory networks without the need to specify higher order
interaction terms or nonlinearities ahead of time.

Additionally, we have data from chromatin immunoprecipitation (ChIP)
binding experiments [\citet{Lee2002}]. Such experiments use antibodies
to isolate specific DNA sequences which are bound by a TF. This
information can be used to discover potential binding locations for
particular transcription factors within the genome. The ChIP data can
be considered
``prior information'' that one may wish to make use of when
investigating gene regulatory networks. Given the Bayesian nature of
our approach, we propose a straightforward modification to \texttt
{BART} which incorporates such prior information into our variable
selection procedure.

In Section~\ref{secvarsel} we review some common techniques for
variable selection. We emphasize the limitations of approaches relying
on linear models and highlight variable selection via tree-based
techniques. We provide an overview of the \texttt{BART} algorithm and
its output in Section~\ref{secBART}. In Sections~\ref{subsecincl}
and \ref{subsecnulldistexpl} we introduce how \texttt{BART}
computes variable inclusion proportions and explore the properties of
these proportions. In Section~\ref{subsecpermdist} we develop
procedures for principled variable selection based upon \texttt{BART}
output. In Section~\ref{secprior} we extend the \texttt{BART}
procedure to incorporate prior information about predictor variable
importance. In Section~\ref{secsimulations} we compare our
methodology to alternative variable selection approaches in both linear
and nonlinear simulated data settings. In Section~\ref{secgeneapp}
we apply our \texttt{BART}-based variable selection procedure to the
discovery of gene regulatory networks in budding yeast. Section~\ref
{secdiscussion} concludes with a brief discussion. We note that our
variable selection procedures as well as the ability to incorporate
informed prior information are readily available features in the
\texttt{R} package \texttt{bartMachine} [\citet{Kapelner2013}],
currently available on CRAN.

%s2 #&#
\section{Techniques for variable selection}\label{secvarsel}

%s2.1 #&#
\subsection{Linear methods}
The variable selection problem has been well studied from both the
classical and Bayesian perspective, though most previous work focuses
on the case where the outcome variable is assumed to be a linear
function of the available covariates. Stepwise regression [\citet{Hoc76}] is a common approach for variable selection from a large set
of possible predictor variables. Best subsets regression [\citet{Mil02}]
can also be employed, although this option becomes too
computationally burdensome as $K$ becomes large. Other popular linear
variable selection methods are lasso regression [\citet{Tibshirani1996}]
and the elastic net [\citet{Zou2005}]. Both of these
approaches enforce sparsity on the subset of selected covariates by
imposing penalties on nonzero coefficients. \citet{Park2008} and
\citet{Han09} provide Bayesian treatments of lasso regression.

Perhaps the most popular Bayesian variable selection strategies are
based on linear regression with a ``spike-and-slab'' prior distribution
on the regression coefficients. Initially proposed by \citet
{Mitchell1988}, who used a mixture prior of a point mass at zero and a
uniform slab, \citet{George1993} went on to use a mixture-of-normals
prior, for which a Markov chain Monte Carlo stochastic search of the
posterior could be easily implemented. Eventually, most applications
gravitated toward a limiting form of the normal mixture with a
degenerate point mass at zero. More recent work involving
spike-and-slab models has been developed in \citet{Ishwaran2005},
\citet{Li2010b}, \citet{Hans2007}, \citet
{Botollo2010}, \citet{Stingo2011}, and \citet{Rockova2013}.
In these approaches, variable selection is based on the posterior
probability that each predictor variable is in the slab distribution,
and sparsity can be enforced by employing a prior that strongly favors
the spike distribution at zero.

%s2.2 #&#
\subsection{Tree-based methods}

Each of the aforementioned approaches assumes that the response
variable is a linear function of the predictor variables. A major
drawback of linear models, both in the frequentist and Bayesian
paradigms, is that they are ill-equipped to handle complex, nonlinear
relationships between the predictors and response. Nonlinearities and
interactions, which are seldom known with certainty, must be specified
in advance by the researcher. In the case where the model is
misspecified, incorrect variables may be included and correct variables
excluded.

As an alternative, we consider nonparametric methods which are flexible
enough to fit a wide array of functional forms. We focus on tree-based
methods, examples of which include random forests [\texttt{RF}, \citet{Bre01}], stochastic gradient boosting [\citet{Friedman2002}], \texttt
{BART}, and dynamic trees [\texttt{DT}, \citet{Taddy2011}]. Compared
with linear models, these procedures are better able to approximate
complicated response surfaces but are ``black-boxes'' in the sense that
they offer less insight into how specific predictor variables relate to
the response variable.

Tree-based variable selection makes use of the internals of the
decision tree structure which we briefly outline. All observations
begin in a single root node. The root node's splitting rule is chosen
and consists of a splitting variable $\mathbf{x}_k$ and a split point
$c$. The observations in the root node are then split into two groups
based on whether $\mathbf{x}_k \geq c$ or $\mathbf{x}_k < c$. These
two groups become a right daughter node and a left daughter node,
respectively. Within each of these two nodes, additional binary splits
can be chosen.

Existing tree-based methods for variable selection focus on the set of
splitting variables within the trees. For example, \citet
{Gramacy2013} develop a backward stepwise variable selection procedure
for \texttt{DT} by considering the average reduction in posterior
predictive uncertainty within all nodes that use a particular predictor
as the splitting variable. Also, the splitting variables in \texttt
{RF} can be used to develop variable selection approaches. For
instance, one can consider the reduction in sum of square errors (node
impurity in classification problems) associated with a particular
predictor. Additionally, \citet{Diaz-Uriarte2006} consider
reduction in out-of-bag mean square error associated with each
predictor to develop a backward stepwise selection procedure.

We too consider the splitting variables for \texttt{BART} in
developing our method, but our approach differs from the previously
mentioned work in two aspects. First, we do not propose a backward
stepwise selection, but rather develop a permutation-based inferential
approach. Second, we do not consider the overall improvement to fit
provided by each predictor variable, but instead consider how often a
particular predictor appears in a \texttt{BART} model. While simple,
this metric shows promising performance for variable selection using
\texttt{BART}.

%s3 #&#
\section{Calibrating \texttt{BART} output for variable selection}\label{secvarselection}

%s3.1 #&#
\subsection{Review of Bayesian Additive Regression Trees}\label{secBART}

\texttt{BART} is a Bayesian ensemble approach for modeling the unknown
relationship between a vector of observed responses $\by$ and a set of
predictor variables $\mathbf{x}_1, \ldots, \mathbf{x}_K$ without
assuming any parametric functional form for the relationship. The key
idea behind \texttt{BART} is to model the regression function by a sum
of regression trees with homoskedastic normal additive noise,
%
%
%e1 #&#
\begin{equation}
\label{eqbart1} \by= \sum
_{i = 1}^{m} {\mathcal
T}_i (\mathbf{x}_1,\ldots,\mathbf{x}_K) +
\bolds{\errorrv}, \qquad\bolds{\errorrv}\sim\mathcal{N}_{n} \bigl(\0,
\sigma^2\mathbf{I}_n \bigr). \label{BARTeq}
\end{equation}

Here, each ${\mathcal T}_i (\mathbf{x}_1,\ldots,\mathbf{x}_K)$ is a
regression tree that partitions the predictor space based on the values
of the predictor variables. Observations with similar values of the
predictor variables are modeled as having a similar predicted response
$\hat{\by}$.

Each regression tree ${\mathcal T}_i$ consists of two components: a
tree structure ${\mathcal T}_i$ and a set of terminal node parameters
$\bmu_i$. The tree ${\mathcal T}_i$ partitions each observation into a
set of $B_i$ terminal nodes based on the splitting rules contained in
the tree. The terminal nodes are parameterized by $\bmu_i =\{\mu
_{i1},\ldots,\mu_{iB_i}\}$ such that each observation contained
within terminal node $b$ is assigned the same response value of $\mu
_{ib}$. Regression trees yield a flexible model that can capture
nonlinearities and interaction effects in the unknown regression function.

As seen in equation~(\ref{BARTeq}), the response vector $\by$ is
modeled by the sum of $m$ regression trees. For each observation, the
predicted response $\hat{y}_j$ is the sum of the terminal node
parameters $\mu_{ib}$ for that observation $j$ from each tree
${\mathcal T}_i$. Compared to a single tree, the sum of trees allows
for easier modeling of additive effects [\citet{Chipman10}]. The
residual variance $\sigma^2$ is considered a global parameter shared
by all observations.

In this fully Bayesian approach, prior distributions must also be
specified for all unknown parameters, which are the full set of tree
structures and terminal node parameters $({\mathcal T}_i, \bmu_i)$,
as well as the residual variance $\sigma^2$. The prior distributions
for $({\mathcal T}_i, \bmu_i)$ are specified to give a strong
preference to small simple trees with modest variation of the terminal
node parameter values, thereby limiting the impact on the model fit of
any one tree. The result is that \texttt{BART} consists of an ensemble
of ``weak learners,'' each contributing to the approximation of the
unknown response function in a small and distinct fashion. The prior
for $\sigma^2$ is the inverse chi-square distribution with
hyperparameters chosen based on an estimate of the residual standard
deviation of the data.

The number of trees $m$ in the ensemble is considered to be a
prespecified hyperparameter. The usual goal of \texttt{BART} is
predictive performance, in which case a large value of $m$ allows for
increased flexibility when fitting a complicated response surface,
thereby improving predictive performance. However, \citet
{Chipman10} recommend using a smaller value of $m$ for the purposes of
variable selection (we default to $m=20$). When the number of trees in
the ensemble is smaller, there are fewer opportunities for predictor
variables to appear in the model and so they must compete with each
other to be included. However, if $m$ is too small, the Gibbs sampler
in \texttt{BART} becomes trapped in local modes more often, which can
destabilize the results of the estimation procedure [\citet{Chipman1998a}]. Also, there is not enough flexibility in the model to
fit a variety of complicated functions. However, when the number of
trees becomes too large, there is opportunity for unimportant variables
to enter the model without impacting the overall model fit, thereby
making variable selection more challenging.

Our explorations have shown that $m=20$ represents a good compromise,
although similar choices of $m$ should not impact results. Under the
sparse data settings we will examine in Sections~\ref{secsimulations}
and \ref{secgeneapp}, we show that this medium level of $m$ aids the
selection of important predictor variables even when the number of
predictor variables is relatively large.

It is also worth noting that in the default \texttt{BART} formulation,
each predictor variable $\mathbf{x}_k$ has an equal {a~priori}
chance of being chosen as a splitting variable for each tree in the
ensemble. However, in many applications, we may have real prior
information that suggests the importance of particular predictor
variables. In Section~\ref{secprior}, we will extend the \texttt
{BART} procedure to incorporate prior information about specific
predictor variables, which will be used to aid in discovering the yeast
gene regulatory network in Section~\ref{secgeneapp}.

The full posterior distribution for the \texttt{BART} model is
estimated using Markov chain Monte Carlo methods. Specifically, a Gibbs
sampler [\citet{GemGem84}] is used to iteratively sample from the
conditional posterior distribution of each set of parameters. Most of
these conditional posterior distributions are standard, though a
Metropolis--Hastings step [\citet{Has70}] is needed to alter the tree
structures ${\mathcal T}_i$. Details are given in \citet
{Chipman10} and \citet{Kapelner2013}.

%s3.2 #&#
\subsection{\texttt{BART} variable inclusion proportions} \label{subsecincl}

The primary output from \texttt{BART} is a set of predicted values
$\hat{\by}$ for the response variable $\by$. Although these
predicted values $\hat{\by}$ serve to describe the overall fit of the
model, they are not directly useful for evaluating the relative
importance of each predictor variable in order to select a subset of
predictor variables. For this purpose, \citet{Chipman10} begin
exploring the
``variable inclusion proportions'' of each predictor variable. We
extend their exploration into a principled method.

Across all $m$ trees in the ensemble (\ref{BARTeq}), we
examine the set of predictor variables used for each splitting rule in
each tree. Within each posterior Gibbs sample, we can compute the
proportion of times that a split using $\mathbf{x}_k$ as a splitting
variable appears among all splitting variables in the ensemble. Since
the output of \texttt{BART} consists of many posterior samples, we
estimate the \textit{variable inclusion proportion} $p_k$ as the
posterior mean of the these proportions across all of the posterior samples.

Intuitively, a large variable inclusion proportion $p_k$ is suggestive
of a predictor variable $\mathbf{x}_k$ being an important driver of
the response. \citet{Chipman10} suggest using $\bp= (p_1,\ldots,p_K)$ to
rank variables $\mathbf{x}_1, \ldots, \mathbf{x}_K$ in terms of
relative importance. These variable inclusion proportions naturally
build in some amount of multiplicity control since the $p_k$'s have a
fixed budget (in that they must sum to one) and that budget will become
more restrictive as the number of predictor variables increases.

However, each variable inclusion proportion $p_k$ cannot be interpreted
as a posterior probability that the predictor variable $\mathbf{x}_k$
has a ``real effect,'' defined as the impact of some linear or
nonlinear association, on the response variable. This motivates the
primary question being addressed by this paper: \textit{how large does
the variable inclusion proportion $p_k$ have to be in order to select
predictor variable $\mathbf{x}_k$ as an important variable}?

As a preliminary study, we evaluate the behavior of the variable
inclusion proportions in a ``null'' data setting, where we have a set of
$K$ predictor variables $\mathbf{x}_k$ that are all unrelated to the
outcome variable $\by$. Specifically, we generate each response
variable $y_i$ and each predictor variable $x_{ik}$ independently from
a standard normal distribution. In this null setting, one might expect
that \texttt{BART} would choose among the predictor variables
uniformly at random when adding variables to the ensemble of trees
[equation~(\ref{BARTeq})]. In this scenario, each variable inclusion
proportion would then be close to the inverse of the number of
predictor variables, that is, $p_k \approx1/K$ for all $k$.

\begin{figure}%[b]

\includegraphics{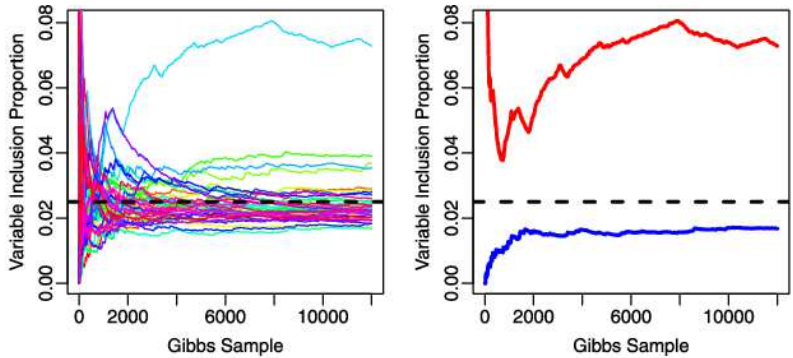}

\caption{Variable inclusion proportions from \texttt{BART} model in
null setting where each predictor variable is unrelated to the response
variable. (Left)~variable inclusion proportions for all $K=40$
predictor variables over 12,000 Gibbs samples. (Right)~tracking
of the maximum and minimum of the variable inclusion proportions.}\label{fignullinclusionprops}
\end{figure}

However, we have found empirically that in this scenario the variable
inclusion proportions do not approach $1/K$ for all predictor
variables. As an example, Figure~\ref{fignullinclusionprops} gives
the variable inclusion proportions from a null simulation with $n=250$
observations and $K=40$ predictor variables, all of which are unrelated
to the response variable $\by$.

%
%f1 #&#

In this setting, the variable inclusion proportions do not converge to
$1/40 =0.025$. As seen in Figure~\ref{fignullinclusionprops}, some
variable inclusion proportions remain substantially larger than $1/K$
and some are substantially smaller. We observed this same phenomenon
with different levels of noise in the response variable.

%s3.3 #&#
\subsection{Further exploration of null simulation}\label{subsecnulldistexpl}

We hypothesize that the variation between $p_k$'s in Figure~\ref
{fignullinclusionprops} can stem from two causes. First, even though
the response and predictors were generated independently, they will
still exhibit some random association. \texttt{BART} may be fitting
noise, or
``chance-capitalizing;'' given its nonparametric flexibility, \texttt
{BART} could be fitting to perceived nonlinear associations that are
actually just noise. Second, there might be inherent variation in the
\texttt{BART} estimation procedure itself, possibly due to the Gibbs
sampler getting stuck in a local maximum.

Thus, we consider an experiment to explore the source of this variation
among the $p_k$'s. We generate 100 data sets under the same setting as
that in Figure~\ref{fignullinclusionprops}. Within each data set,
we run \texttt{BART} 50 times with different initial values for the
model parameters randomly drawn from the respective prior
distributions. Let $p_{ijk}$ denote the variable inclusion proportion
for the $i$th data set, $j$th \texttt{BART} run, and the
$k$th predictor variable. We then consider the decomposition into
three nested variances listed in Table~\ref{tabnestedvariances}.
Note that we use standard deviations in our illustration that follows.

%
%t1 #&#
\begin{table}[b]
\tabcolsep=0pt
\caption{The three nested variances}\label{tabnestedvariances}
\begin{tabular*}{\tablewidth}{@{\extracolsep{\fill}}@{}ll@{}}
\hline
$s^2_{ik} = \frac{1}{50}\sum_{j=1}^{50}(p_{ijk} - \pbar _{i\cdot k})^2$
& The variability of \texttt{BART} estimation for a particular predictor\\
&  $k$ in a particular data set $i$
\\[3pt]
$s^2_{k} = \frac{1}{100}\sum_{i=1}^{100}( \pbar_{i\cdot k}- \pbar_{\cdot\cdot k})^2$
& The variability due to chance capitalization of the \texttt{BART} \\
& procedure for predictor $k$ across data sets
\\[3pt]
$s^2 = \frac{1}{40}\sum_{k=1}^{40}(\pbar_{\cdot\cdot k} -\pbar_{\cdot\cdot\cdot})^2$
& The variability across predictors
\\
\hline
\end{tabular*}
\end{table}

First consider what may happen if the source of Figure's~\ref{fignullinclusionprops} observed pathology is purely due to
\texttt{BART}'s Gibbs sampler getting stuck in different local
posterior modes. On the first run for the first data set, \texttt
{BART} would fall into a local mode where some predictors are naturally
more important than others and, hence, the $p_{11k}$'s would be
unequal. In the same data set, second run, \texttt{BART} might fall
into a different local mode where the $p_{12k}$'s are unequal, but in a
way that is different from the first run's $p_{11k}$'s. This type of
process would occur over all 50 runs. Thus, the $s_{1k}$, the standard
deviation of $p_{ijk}$ over runs of \texttt{BART} on the first data
set, would be large. Note that if there is no chance capitalization or
overfitting, there should be no reason that averages of the
proportions, the $\pbar_{1 \cdot k}$'s, should be different from $1/K$
over repeated runs. Then, when the second data set is introduced,
\texttt{BART} will continue to get stuck in different local posterior
modes and the $s_{2k}$'s should be large, but the $\pbar_{2 \cdot
k}$'s should be near $1/K$. Hence, over all of the data sets, $\pbar
_{i \cdot k}$'s should be approximately $1/K$, implying that the
$s_k$'s should be small. In sum, \texttt{BART} getting stuck in local
modes suggests large $s_{ik}$'s and small $s_k$'s.

Next consider what may happen if the source of Figure's~\ref{fignullinclusionprops} observed pathology is purely due to
\texttt{BART} chance-capitalizing on noise. On the first data set,
over each run, \texttt{BART} does not get stuck in local modes and,
therefore, the $p_{i1k}$'s across runs would be fairly stable. Hence,
the $s_{1k}$'s would be small. However, in each of the runs, \texttt
{BART} overfits in the \textit{same way} for each data set. For
example, perhaps \texttt{BART} perceives an association between $x_1$
and $y$ on the first data set. Hence, the $p_{1j1}$'s would be larger
than $1/K$ on all restarts (\texttt{BART} would select $x_1$ as a
splitting rule often due to the perceived association) and, thus,
$\pbar_{1 \cdot1} > 1/K$. Then, in the second data set, \texttt
{BART} may perceive an association between $x_3$ and $y$, resulting in
$p_{2j3}$'s being larger on all runs ($\pbar_{2 \cdot3} > 1/K$).
Thus, \texttt{BART} overfitting is indicated by small $s_{ik}$'s and
large $s_k$'s.

%
%f2 #&#
\begin{figure}[b]

\includegraphics{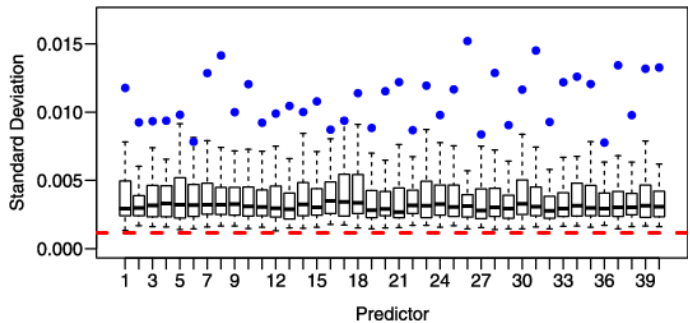}

\caption{The boxplots represent the distribution of $s_{ik}$ for each
predictor. The circles represent the values of $s_k$ and the dashed
line corresponds to $s$. Note that the results are reported as standard
deviations and points in the boxplots beyond the whiskers are omitted.}\label{figanova}
\end{figure}

Figure~\ref{figanova} illustrates the results of the simulations.
Both sources of variation appear, but for all predictors, the average
$s_{ik}$ is significantly smaller than the $s_k$. This finding suggests
that within a particular data set, \texttt{BART} is
chance-capitalizing and overfitting to the noise, which prevents the
$p_k$'s from converging to $1 / K$.\footnote{We also considered this
experiment with orthogonalized predictors (not shown). This reduces the
$s_k$'s (chance capitalization) in Figure~\ref{figanova} slightly,
but the $s_k$'s are still larger than the average $s_{ik}$'s. Hence,
even if there is no \textit{linear} correlation between the predictors
and the response, \texttt{BART} is capitalizing on nonlinear associations.}

Also note the overall average inclusion proportion $\pbar_{\cdot\cdot
\cdot}$ is $0.025 = 1/K$, so across data sets and \texttt{BART} runs
the variable inclusion proportions are correct on average. Further, the
standard deviation across predictors $s$ is small. This implies that
the $\pbar_{\cdot\cdot k}$'s are approximately $1/K$ as well, which
indicates there is no systematic favoring of different covariates once
the effect of overfitting by data set and remaining in local modes by
run is averaged out.

We believe this experiment demonstrates that there is a large degree of
chance capitalization present in the variable inclusion proportions in
the ``null'' model. This implies that it is not possible to decide on
an appropriate threshold for the $p_k$'s when selecting a subset of
important predictor variables in real data settings. Further, the
chance capitalization is idiosyncratic for any data set, making it
challenging to pose a simple parametric model for the behavior in
Figure~\ref{fignullinclusionprops} that would be useful in
practice. This motivates our nonparametric approach to establishing
thresholds for the variable inclusion proportions based on permutations
of the response variable $\by$.

As noted above, there is some variability in the $p_k$'s between
\texttt{BART} runs from different starting points. We found that
averaging over results from five repetitions of the \texttt{BART}
algorithm from different starting points was sufficient to provide
stable estimates of the variable inclusion proportions and use these
averaged values as our variable inclusion proportions for the remainder
of the article.

%s3.4 #&#
\subsection{Variable inclusion proportions under permuted responses}\label{subsecpermdist}

We now address our key question: \textit{how large does the variable
inclusion frequency $p_k$ have to be in order to select predictor
variable $\mathbf{x}_k$}? To determine an appropriate selection
threshold, we employ a permutation-based approach to generate a null
distribution for the variable inclusion proportions $\bp= (p_1,\ldots,p_K)$.

Specifically, we create $P$ permutations of the response vector: $\by
^*_1, \by^*_2, \ldots,\by^*_P$. For each of these permuted response
vectors $\by^*_p$, we run the \texttt{BART} model using $\by^*_p$ as
the response and the original $\mathbf{x}_1, \ldots, \mathbf{x}_K$
as predictor variables. This permutation strategy preserves possible
dependencies among the predictor variables while removing any
dependency between the predictor variables and the response variable.

We retain the variable inclusion proportions estimated from the \texttt
{BART} run using each permuted response $\by^*_p$. We use the notation
$p^*_{k,p}$ for the variable inclusion proportion from \texttt{BART}
for predictor $\mathbf{x}_k$ from the $p$th permuted response, and we
use the notation $\bp^*_{p}$ for the vector of all variable inclusion
proportions from the $p$th permuted response. We use the variable
inclusion proportions $\bp^*_1, \bp^*_2, \ldots,\bp^*_P$ across all
$P$ permutations as the null distribution for our variable inclusion
proportions $\bp$ from the real (unpermuted) response~$\by$.

The remaining issue is selecting an appropriate threshold for predictor
$\mathbf{x}_k$ based on the permutation null distribution $\bp^*_1,
\bp^*_2, \ldots,\bp^*_P$. We will consider three different threshold
strategies that vary in terms of the stringency of their resulting
variable selection procedure.

The first strategy is a {``\emph{local}''} threshold: we calculate a
threshold for each variable inclusion proportion $p_k$ for each
predictor $\mathbf{x}_k$ based only on the permutation null
distribution of $p_k$. Specifically, we take the $1-\alpha$ quantile
of the distribution of $p^*_{k,1}, p^*_{k,2}, \ldots, p^*_{k,P}$ and
only select predictor $\mathbf{x}_k$ if $p_k$ exceeds this $1-\alpha$
quantile.\vspace*{1pt}

The second strategy is a {``\emph{global max}''} threshold: we calculate a
threshold for the variable inclusion proportion $p_k$ for predictor
$\mathbf{x}_k$ based on the maximum across the permutation
distributions of the variable inclusion proportions for all predictor
variables. Specifically, we first calculate $p^*_{\max,p} = \max
\{p^*_{1,p}, p^*_{2,p}, \ldots, p^*_{K,p} \}$, the largest
variable inclusion proportion across all \mbox{predictor} variables in
permutation $p$. We then calculate the $1-\alpha$ quantile of the
distribution of $p^*_{\max,1}, p^*_{\max,2}, \ldots,
p^*_{\max,P}$ and only select\vspace*{1pt} predictor $\mathbf{x}_k$ if $p_k$
exceeds this $1-\alpha$ quantile.

The first ``local'' strategy and the second ``global max'' strategy are
opposite extremes in terms of the stringency of the resulting variable
selection. The local strategy is least stringent since the variable
inclusion proportion $p_k$ for predictor $\mathbf{x}_k$ needs to only
be extreme within its own permutation distribution in order to be
selected. The global maximum strategy is most stringent since the
variable inclusion proportion $p_k$ for predictor $\mathbf{x}_k$ must
be extreme relative to the permutation distribution across all
predictor variables in order to be selected.

We consider a third strategy that is also global across predictor
variables, but is less stringent than the global max strategy. The
third {``\emph{global SE}''} strategy uses the mean and standard deviation
from the permutation distribution of each variable inclusion proportion
$p_k$ to create a global threshold for all predictor variables.
Specifically, letting $m_k$ and $s_k$ be the mean and standard
deviation of variable inclusion proportion $p^*_k$ for predictor
$\mathbf{x}_k$ across all permutations, we calculate
\[
C^* = \inf_{C\in\mathbb{R}^+} \Biggl\{\forall k,
\frac{1}{P}\sum_{p=1}^P\mathbb{I}
\bigl(p^*_{k,p} \leq m_k + C \cdot s_k \bigr) >
1 - \alpha\Biggr\}.
\]

The value $C^*$ is the smallest global multiplier that gives
simultaneous $1-\alpha$ coverage across the permutation distributions
of $p_k$ for all predictor variables. The predictor $\mathbf{x}_k$ is
then only selected if $p_k > m_k + C^*\cdot s_k$. This third strategy
is a compromise between the local permutation distribution for variable
$k$ (by incorporating each mean $m_k$ and standard deviation $s_k$) and
the global permutation distributions of the other predictor variables
(through $C^*$). We outline all three thresholding procedures in more
detail in the \hyperref[thresholdalgorithms]{Appendix}.

%
%f3 #&#
\begin{figure}[b]

\includegraphics{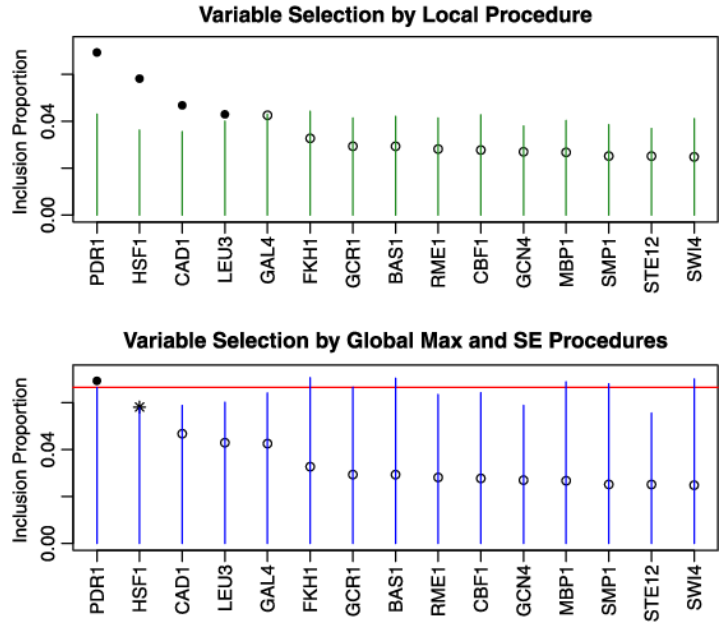}

\caption{The fifteen largest variable inclusion proportions from
\texttt{BART} implemented on the yeast gene \texttt{YAL004W} with
$\alpha=0.05$. (Top)~the tips of the green bands are the ``local''
thresholds of our first variable selection strategy. Solid dots are
selected predictor variables, whereas hollow dots are unselected
predictor variables. (Bottom)~the red line is the threshold from
our second ``global max'' strategy. The tips of the blue bands are the
thresholds from our ``global SE'' strategy. The one solid dot is the
predictor selected by both strategies. The star is the additional
predictor variable selected by only the global SE strategy. The hollow
dots are unselected predictor variables.}\label{figselectionex}
\end{figure}

As an example of these three thresholding strategies, we provide a
brief preview of our application to the yeast gene regulatory network
in Section~\ref{secgeneapp}. In that application, the response
variable $\by$ consists of the expression measures for a particular
gene across approximately 300 conditions and the predictor variables
are the expression values for approximately 40 transcription factors in
those same 300 conditions.

In Figure~\ref{figselectionex}, we give the fifteen predictor
variables with the largest variable inclusion proportions from the
\texttt{BART} model implemented on the data for a particular yeast
gene \texttt{YAL004W}. In the top plot, we see the different ``local''
thresholds for each predictor variable. Four of the predictor variables
had variable inclusion proportions $p_k$ that exceeded their local
threshold and were selected under this first strategy. In the bottom
plot, we see the single ``global max'' threshold for all predictor
variables as well as the different ``global SE'' thresholds for each
predictor variable. Two of the predictor variables had variable
inclusion proportions $p_k$ that exceeded their global SE thresholds,
whereas only one predictor variable exceeded the global max threshold.

This example illustrates that our three threshold strategies can differ
substantially in terms of the stringency of the resulting variable
selection. Depending on our prior expectations about the sparsity in
our predictor variables, we may prefer the high stringency of the
global max strategy, the low stringency of the local strategy, or the
intermediary global SE strategy.

In practice, it may be difficult to know {a~priori} the level of
stringency that is desired for a real data application. Thus, we
propose a \emph{cross-validation strategy} for deciding between our
three thresholding strategies for variable selection. Using $k$-fold
cross-validation, the available observations can be partitioned into
training and holdout subsets. For each partition, we can implement all
three thresholding strategies on the training subset of the data and
use the thresholding strategy with the smallest prediction error across
the holdout subsets. We call this procedure ``\texttt{BART}-Best'' and
provide implementation details in the \hyperref[thresholdalgorithms]{Appendix}.

Our permutation-based approach for variable selection does not require
any additional assumptions beyond those of the \texttt{BART} model.
Once again, the sum-of-trees plus normal errors is a flexible
assumption that should perform well across a wide range of data
settings, especially relative to methods that make stronger parametric
demands. Also, it is important to note that we view each of the
strategies described in this section as a procedure for variable
selection based on well-founded statistical principles, but do not
actually associate any particular formal hypothesis testing with our
approach. Finally, a disadvantage of our permutation-based proposal is
the computational cost of running \texttt{BART} on a large set of
permuted response variables $\by^*$. However, it should be noted that
the permuted response vector runs can be computed in parallel on
multiple cores when such resources are available.

%s3.5 #&#
\subsection{Real prior information in \texttt{BART}-based variable selection} \label{secprior}

Most variable selection approaches do not allow for {a~priori}
preferences for particular predictor variables. However, in many
applications, there may be available prior information that suggests
particular predictor variables may be more valuable than others.

As an example, the yeast regulatory data in Section~\ref{secgeneapp}
consist of expression measures $\by_g$ for a particular gene $g$ as
the response variable with predictor variables $\mathbf{x}_k$ being
the expression values for $\approx$40 transcription factors. In
addition to the expression data, we also have an accompanying
ChIP-binding data set [\citet{Lee2002}] that indicates for each gene
$g$ which of the $\approx$40 transcription factors are likely to bind
near that gene. We can view these ChIP-binding measures as prior
probabilities that particular predictor variables $\mathbf{x}_k$ will
be important for the response variable $\by$.

The most natural way to give prior preference to particular variables
in \texttt{BART} is to alter the prior on the splitting rules. As
mentioned in Section~\ref{secBART}, by default each predictor
$\mathbf{x}_k$ has an equal {a~priori} chance of being chosen as a
splitting rule for each tree branch in the \texttt{BART} ensemble. We
propose altering the prior of the standard \texttt{BART}
implementation so that when randomly selecting a particular predictor
variable for a splitting rule, more weight is given to the predictor
variables that have a higher prior probability of being important.
Additionally, the prior on the tree structure, which is needed for the
Metropolis--Hastings ratio computation, is appropriately adjusted. This
strategy has some precedent, as \citet{Chipman1998a} discuss
nonuniform criteria for splitting rules in the context of an earlier
Bayesian Classification and Regression Tree implementation. Note that
when employing one of the strategies discussed in Section~\ref
{subsecpermdist}, the prior is reset to discrete uniform when
generating the permutation distribution, as it is assumed that there is
no relationship between the predictors and the response.

In Section~\ref{secsim-prior} we present a simulation-based
evaluation of the effects on correct variable selection when an
informed prior distribution is either correctly specified, giving
additional weight to the predictor variables with true influence on the
response, or incorrectly specified, giving additional weight to
predictor variables that are unrelated to the response. Before our
simulation study of the effects of prior information, we first present
an extensive simulation study that compares our \texttt{BART}-based
variable selection procedure to several other approaches.

%s4 #&#
\section{Simulation evaluation of \texttt{BART}-based variable selection}\label{secsimulations}

We use a variety of simulated data settings to evaluate the ability of
our \texttt{BART}-based procedure to select the subset of predictor
variables that have a true influence on a response variable. We examine
settings where the response is a linear function of the predictor
variables in Section~\ref{secsim-linear} as well as settings where
the response is a nonlinear function of the predictor variables in
Section~\ref{secsim-nonlinear}. We also examine the effects of
correctly versus incorrectly specified informed prior distributions in
Section~\ref{secsim-prior}. For each simulated data setting, we will
compare the performance of several different variable selection approaches:

\begin{longlist}[(1)]
\item[(1)]\texttt{BART}-\textit{based variable selection}: As outlined in
Section~\ref{secvarselection}, we use the variable inclusion
proportions from \texttt{BART} to rank and select predictor variables.
We will evaluate the performance of the three proposed thresholding
strategies as well as ``\texttt{BART}-Best,'' the (five-fold)
cross-validation strategy for choosing among our thresholding
strategies. In each case, we set $\alpha= 0.05$ and the number of
trees $m$ is set to 20. Default settings from \citet{Chipman10}
are used for all other hyperparameters. The variable selection
procedures are implemented in the \texttt{R} package \texttt
{bartMachine} [\citet{Kapelner2013}].

\item[(2)]\textit{Stepwise regression}: Backward stepwise regression using
the \texttt{stepAIC} function in \texttt{R}.\footnote{We also
considered forward stepwise regression but found that backward stepwise
regression performed better in our simulated data settings.}

\item[(3)]\textit{Lasso regression}: Regression with a lasso (L1) penalty
can be used for variable selection by selecting the subset of variables
with nonzero coefficient estimates. For this procedure, an additional
penalty parameter $\lambda$ must be specified, which controls the
amount of shrinkage toward zero in the coefficients. We use the {\tt
glmnet} package in \texttt{R} [\citet{Friedman2010}], which uses
ten-fold cross-validation to select the value of the penalty parameter
$\lambda$.

\item[(4)]\textit{Random forests} ({\texttt{RF}}): Similarly to \texttt
{BART}, \texttt{RF} must be adapted to the task of variable
selection.\footnote{Existing variable selection implementations for
\texttt{RF} from \citet{Diaz-Uriarte2006} and \citet
{Deng2012} are not implemented for regression problems to the best of
our knowledge.} The {\tt randomForest} package in \texttt{R} [\citet{Liaw2002}] produces an ``importance score'' for each predictor
variable: the change in out-of-bag mean square error when that
predictor is not allowed to contribute to the model. \citet
{Breiman2013} suggest selecting only variables where the importance
score exceeds the $1-\alpha$ quantile of a standard normal
distribution. We follow their approach and further suggest a new
approach: using the Bonferroni-corrected $(1-\alpha)/p$ quantile of a
standard normal distribution. We employ a five-fold cross-validation
approach to pick the best of these two thresholding strategies in each
simulated data setting and let $\alpha=0.05$. Default parameter
settings for \texttt{RF} are used.

\item[(5)]\textit{Dynamic trees} ({\texttt{DT}}): \citet{Gramacy2013}
introduce a backward variable selection procedure for \texttt{DT}. For
each predictor, the authors compute the average reduction in posterior
predictive uncertainty across all nodes using the given predictor as a
splitting variable. The authors then propose a relevance probability,
which is the proportion of posterior samples in which the reduction in
predictive uncertainty is positive. Variables are deselected if their
relevance probability does not exceed a certain threshold. After
removing variables, the procedure is repeated until the log-Bayes
factor of the larger model over the smaller model is positive,
suggesting a preference for the larger model. We construct \texttt{DT}
using the \texttt{R} package \texttt{dynaTree} [\citet{Taddy2011}]
with 5000 particles and a constant leaf model. We employ the default
relevance threshold suggested by the authors of 0.50.

\item[(6)]\textit{Spike-and-slab regression} ({\texttt{Spike-slab}}): We
employ the spike-and-slab regression procedure outlined in \citet
{Ishwaran2005} and \citet{Ishwaran2010}. The procedure first fits
a spike-and-slab regression model and then performs variable selection
via the generalized elastic net. Variables with nonzero coefficients
are considered relevant. The method is applicable to both high- and
low-dimensional problems, as in the high-dimensional setting, a
filtering of the variables is first performed for dimension reduction.
The procedure is implemented in the \texttt{R} package \texttt{Spikeslab} [\citet{Ishwaran2013}].
\end{longlist}

Each of the above methods will be compared on the ability to select
``useful'' predictor variables, the subset of predictor variables that
truly affect the response variable. We can quantify this performance by
tabulating the number of true positive (TP) selections, false positive
(FP) selections, true negative (TN) selections, and false negative (FN)
selections. The \textit{precision} of a variable selection method is
the proportion of truly useful variables among all predictor variables
that are selected,
%
%
%e2 #&#
\begin{equation}
\label{eqprecision} \mathrm{Precision} = \frac{\mathrm{TP}}{\mathrm{TP}+{\mathrm{FP}}}.
\end{equation}

The \textit{recall} of a variable selection method is the proportion
of truly useful variables selected among all truly useful predictor variables,
%
%
%e3 #&#
\begin{equation}
\label{eqrecall} \mathrm{Recall} = \frac{\mathrm{TP}}{\mathrm{TP}+\mathrm{FN}}.
\end{equation}

We can combine the precision and recall together into a single
performance criterion,
%
%
%e4 #&#
\begin{equation}
\label{eqF1score} F_1 = 2\cdot\frac{\mathrm{Precision}\cdot\mathrm
{Recall}}{\mathrm
{Precision}+\mathrm{Recall}},
\end{equation}

\noindent which is the harmonic mean of precision and recall, balancing
a procedure's capability to make necessary identifications with its
ability to avoid including irrelevant predictors. This \textit{$F_1$
measure} is called the
``effectiveness'' by \citet{vanRijsbergen1979} and is used
routinely in information retrieval and categorization problems.

While many variable selection simulations found in the literature rely
on out-of-sample root mean square error (RMSE) to assess performance of
a procedure, we believe the $F_1$ score is a better alternative.
Out-of-sample RMSE inherently overweights recall vis-\`{a}-vis
precision since predictive performance depends more heavily on
including covariates which generate signal. This is especially true for
adaptive learning algorithms.

We chose the balanced\footnote{The $F_1$ measure can be generalized
with different weights on precision and recall.} $F_1$ metric because
we want to demonstrate flexible performance while balancing both recall
and precision. For example, if an investigator is searching for harmful
physiological agents that can affect health outcomes, identifying the
complete set of agents is important (recall). If the investigator is
looking to fund new, potentially expensive research based on
discoveries (as in our application in Section~\ref{secgeneapp}),
avoiding fruitless directions is most important (precision).

%s4.1 #&#
\subsection{Simulation setting 1: Linear relationship}\label{secsim-linear}

We first examine the performance of the various variable selection
approaches in a situation where the \mbox{response} variable is a linear
function of the predictor variables. Specifically, we generate each
predictor vector $\mathbf{x}_j$ from a normal distribution
%
%
%e5 #&#
\begin{equation}
\label{eqmakecovariates} \mathbf{x}_1, \ldots, \mathbf{x}_p
\stackrel{\mathrm{i.i.d.}} {\sim} \mathcal{N}_{n} (\mathbf{0}, \bI),
\end{equation}
and then the response variable $\by$ is generated as
%
%
%e6 #&#
\begin{equation}
\by= \bX\bbeta+ \bolds{\errorrv}, \qquad\bolds{\errorrv}\sim
\mathcal{N}_{n} \bigl(\0, \sigma^2\bI\bigr),
\end{equation}
where $\bbeta=[\mathbf{1}_{p_0},\mathbf{0}_{p-p_0}]^\top$.
In other words, there are $p_0$ predictor variables that are truly
related to the response $\by$, and $p-p_0$ predictor variables that
are spurious. The \emph{sparsity} of a particular data setting is
reflected in the proportion $p_0/p$ of predictor variables that
actually influence the response.

Fifty data sets were generated for each possible combination of the
following different parameter settings: $p \in{}$\{20, 100, 200,
500, 1000\}, $p_0 / p \in{}$\{0.01, 0.05, 0.1, 0.2\} and $\sigma^2\in{}$\{1, 5, 20\}. In each of the 60
possible settings, the sample size was fixed at $n=250$.

\begin{figure}%[t]

\includegraphics{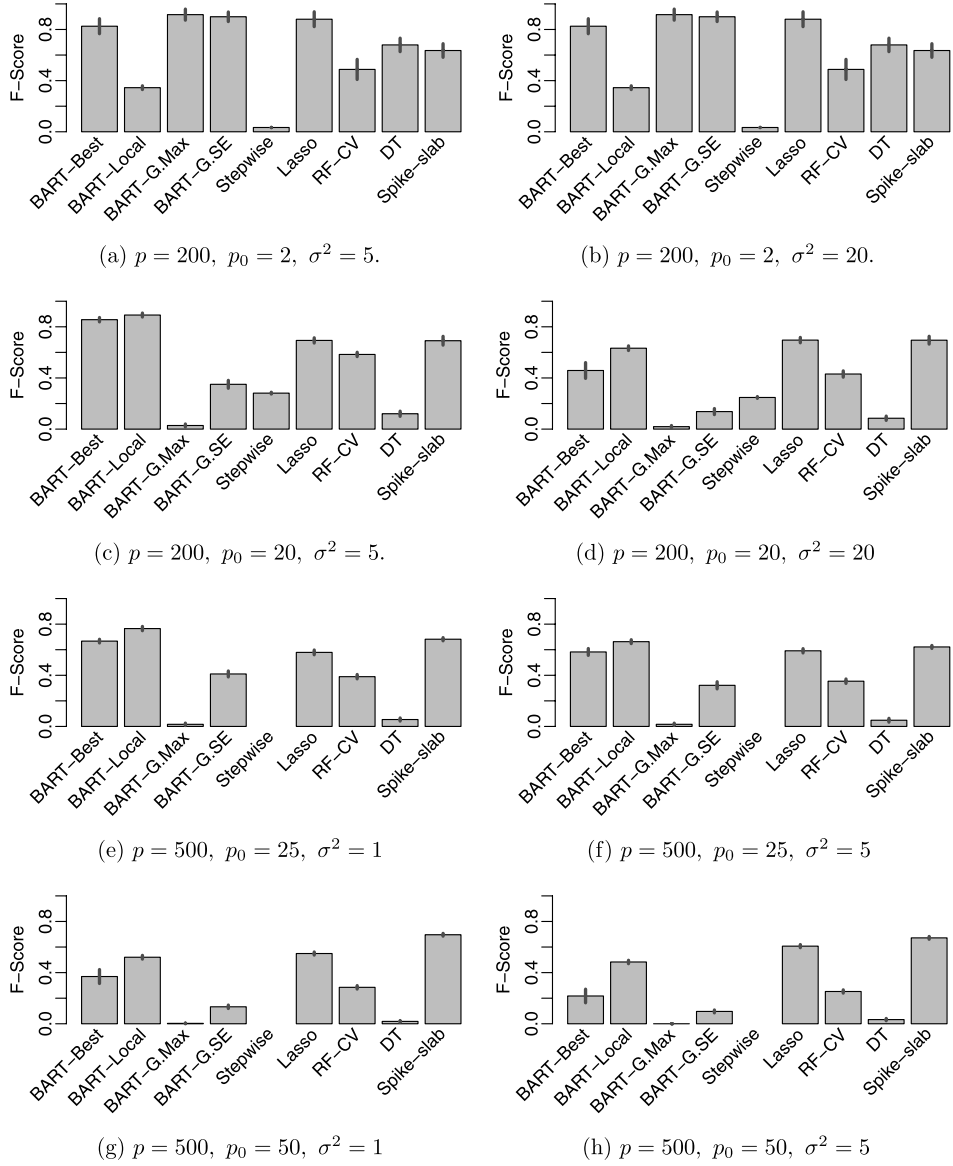}

%a\caption{$p=200,~p_0=2,~\sigma^2=5$.}
%b\caption{$p=200,~p_0=2,~\sigma^2=20$.}
\caption{Average $F_1$ measures for different variable selection
approaches on simulated data under the linear model setting across 50
simulations. The black bars represent 90\% error bars for the average.
Results for $p=200$ and $p=500$ are shown. Within each choice of $p$,
moving down a column shifts from high to low sparsity and moving across
a row shifts from low to high noise.}\label{figF1-linear}\label{figlin1}\label{figlin2}\label{figlin3}\label{figlin4}\label{figlin5}\label{figlin6}\label{figlin7}\label{figlin8}
\end{figure}

Figure~\ref{figF1-linear} gives the $F_1$ performance measure for
each variable selection method for 8 of the 60 simulation settings. We
have chosen to illustrate these simulation results, as they are
representative of our overall findings. Here, higher values of $F_1$
indicate better performance. Complete tables of precision, recall, and
$F_1$ measure values for the simulations shown in Figure~\ref
{figF1-linear} can be found in the supplementary materials [\citet{Bleich2014}].

%
%f4 #&#

We first focus on the comparisons in performance between the four
thresholding strategies for our \texttt{BART}-based variable selection
procedure: our three thresholding strategies plus the \texttt
{BART}-Best cross-validated threshold strategy. First, we consider the
case where $p=200$. In the more sparse settings [Figure~\ref{figlin1}(a)~and~(b)], the more stringent global max and
global SE strategies perform better than the less stringent local
thresholding strategy. However, the local thresholding strategy
performs better in the less sparse settings [Figure~\ref{figlin3}(c)~and~(d)]. The \texttt{BART}-Best procedure with a
cross-validated threshold performs slightly worse than the best of the
three thresholds in each setting, but fares quite well uniformly.
Hence, the cross-validated threshold strategy represents a good choice
when the level of sparsity is not known {a~priori}.

For the settings where $p=500$, the findings are relatively similar.
The local thresholding strategy performs well given the fact that the
data is less sparse. Performance also degrades when moving from the low
noise settings [Figure~\ref{figlin5}(e)~and~(f)] to the
high noise settings [Figure~\ref{figlin6}(g)~and~(h)]. Note
that \texttt{BART}-Best does not perform particularly well in
Figure~\ref{figlin8}(h).

Comparing with the alternative approaches when $p = 200$, we see that
\mbox{\texttt{BART}-}Best performs better than all of the alternatives in the
lower noise, more sparse setting [Figure~\ref{figlin1}(a)] and is
competitive with the lasso in the lower noise, less sparse setting
[Figure~\ref{figlin3}(c)]. \texttt{BART}-Best is competitive with the
lasso in the higher noise, more sparse setting [Figure~\ref{figlin2}(b)]
and beaten by the linear methods in the higher noise, less sparse
setting [Figure~\ref{figlin4}(d)]. When $p=500$, the cross-validated
\texttt{BART} is competitive with the lasso and \texttt{Spike-slab}
and outperforms the nonlinear methods when $p_0 = 25$ [Figure~\ref{figlin5}(e)~and~(f)]. When $p_0 = 50$
[Figure~\ref{figlin7}(g)~and~(h)], the cross-validated \texttt{BART}
performs worse than the lasso and \texttt{Spike-slab}, and has
performance on par with the cross-validated \texttt{RF}.

Overall, the competitive performance of our \texttt{BART}-based
approach is especially impressive since \texttt{BART} does not assume
a linear relationship between the response and predictor variables. One
would expect that stepwise regression, lasso regression, and \texttt
{Spike-slab} would have an advantage since these methods assume a
linear model which matches the data generating process in this setting.
Like \texttt{BART}, \texttt{RF} and \texttt{DT} also do not assume a
linear model, but in most of the cases we examined, our \texttt{BART}-based variable selection procedure performs better than \texttt
{RF} and \texttt{DT}. We note that \texttt{DT} does not perform well
on this simulation, possibly suggesting the need for a cross-validation
procedure to choose appropriate relevance thresholds in different data settings.

Additionally, we briefly address the computational aspect of our four
proposed approaches here by giving an estimate of the runtimes. For
this data with $n=250$ and $p=200$, the three strategies (local, global
max, and global SE) are estimated together in one \texttt{bartMachine}
function in about 90 seconds. The cross-validated \texttt{BART}-Best
procedure takes about 7 minutes.

%s4.2 #&#
\subsection{Simulation setting 2: Nonlinear relationship}\label{secsim-nonlinear}

We next examine the performance of the variable selection methods in a
situation where the response variable is a nonlinear function of the
predictor variables. Specifically, we generate each predictor vector
$\mathbf{x}_j$ from a uniform distribution,
\[
\mathbf{x}_1, \ldots, \mathbf{x}_p \stackrel{
\mathrm{i.i.d.}} {\sim}\mathrm{U}_n (0,1 ),
\]
and then the response variable $\by$ is generated as
%
%
%e7 #&#
%e8 #&#
\begin{eqnarray}\label{eqfriedman}
\by &=& 10\sin{\pi\mathbf{x}_1\mathbf{x}_2}+20(
\mathbf{x}_3-0.5)^2+10\mathbf{x}_4+5
\mathbf{x}_5 + \bolds{\errorrv},
\nonumber\\[-8pt]\\[-8pt]
\eqntext{\bolds{\errorrv}\sim\mathcal{N}_{n} \bigl(\0, \sigma^2\bI\bigr).}
\end{eqnarray}

This nonlinear function from \citet{Friedman1991}, used to showcase
\texttt{BART} in \citet{Chipman10}, is challenging for variable
selection models due to its interactions and nonlinearities. In this
data setting, only the first five predictors truly influence the
response, while any additional predictor variables are spurious.

Fifty data sets were generated for each possible combination of $\sigma
^2\in{}$\{5, 100, 625\} and $p \in{}$\{25, 100, 200, 500,
1000\}. Since the number of relevant predictor variables is
fixed at five, we simulate over a wide range of sparsity values ranging
from $p_0/p = 0.2$ down to $p_0/p = 0.005$. In each data set, the
sample size was fixed at $n=250$.

%
%f5 #&#
\begin{figure}[b]

\includegraphics{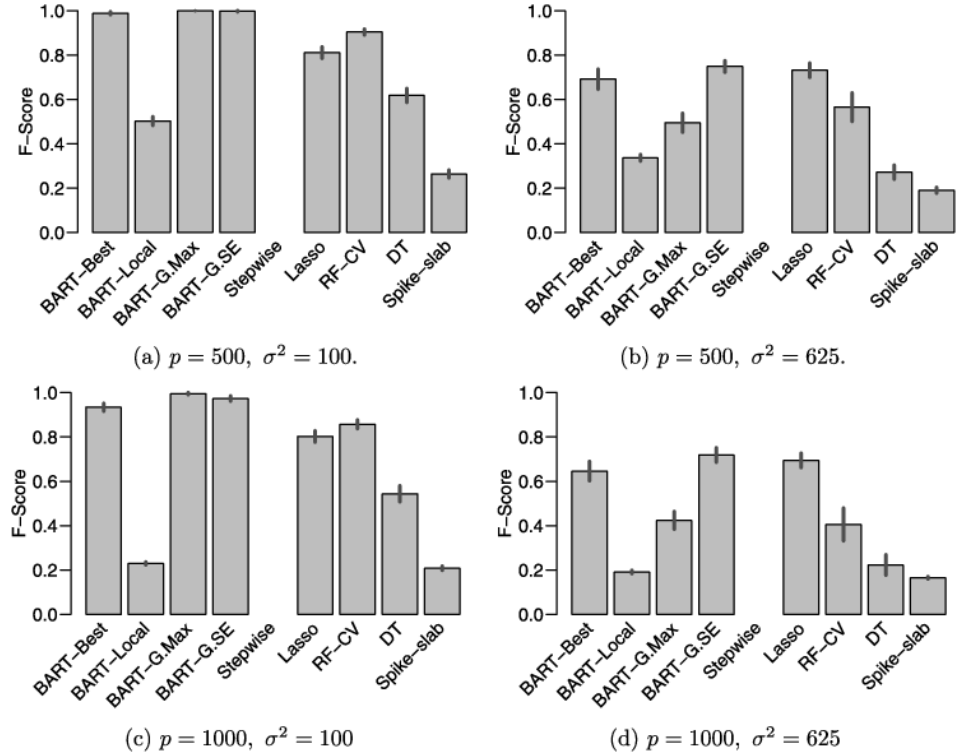}

\caption{Average $F_1$ measures across 50 simulations for different
variable selection approaches on simulated data under the Friedman
model setting. The black bars represent 90\% error bars for the
average. Moving from the top row to the bottom shifts from low to high
dimensionality and moving from the left column to the right shifts from
low to high noise.}\label{figF1-nonlinear}\label{figfr1}\label{figfr2}\label{figfr3}\label{figfr4}
\end{figure}

Figure~\ref{figF1-nonlinear} illustrates the $F_1$ performance
measure for each variable selection method for four of the $
(p,\sigma^2 )$ simulation pairs. We have chosen to illustrate
these simulation results, as they are representative of our overall
findings. Backward stepwise regression via \texttt{stepAIC} could not
be run in these settings where $n < p$ and is excluded from these
comparisons (values in Figure~\ref{figF1-nonlinear} for this
procedure are set to 0). Complete tables of precision, recall, and
$F_1$ measure values for the simulations shown in Figure~\ref
{figF1-nonlinear} are given in our supplementary materials [\citet{Bleich2014}].

Just comparing the four thresholding strategies of our \texttt
{BART}-based procedure, we see that the more stringent selection
criteria have better $F_1$ performance measures in all of these sparse
cases. The cross-validated threshold version of our \texttt{BART}
procedure performs about as well as the best individual threshold in
each case.

Compared to the other variable selection procedures, the
cross-validated \texttt{BART}-Best has the strongest overall
performance. Our cross-validated procedure outperforms \texttt{DT} and
\texttt{RF-CV} in all situations. The assumption of linearity puts the
lasso and \texttt{Spike-slab} at a disadvantage in this nonlinear
setting. \texttt{Spike-slab} does not perform well on this data,
although lasso performs well.\footnote{We note that the lasso's
performance here is unexpectedly high. For this example, lasso is able
to recover the predictors that are interacted within the sine function.
This seems to be an artifact of this particular data generating
process, and we would expect lasso to perform worse on other nonlinear
response functions.} \texttt{BART}-Best and the cross-validated
\texttt{RF} have the best performance in the low noise settings
[Figure~\ref{figfr1}(a)~and~(c)], as they do not assume
linearity. Moving to the high noise settings [Figure~\ref{figfr2}(b)~and~(d)], \texttt{BART} and \texttt{RF} both see a
degradation in performance, and \texttt{BART}-Best and the lasso are
the best performers, followed by the cross-validated~\texttt{RF}.

%s4.3 #&#
\subsection{Simulation setting 3: Linear model with informed priors}\label{secsim-prior}

In the next set of simulations, we explore the impact of incorporating
informed priors into the \texttt{BART} model, as discussed in
Section~\ref{secprior}. We will evaluate the performance of our
\texttt{BART}-based variable selection procedure in cases where the
prior information is correctly specified as well as in cases where the
prior information is incorrectly specified.

We will use the linear model in Section~\ref{secsim-linear} as our
data generating process. We will consider a specific case of the scheme
outlined in Section~\ref{secprior} where particular subsets of
predictor variables are given twice as much weight as the rest of the
predictor variables. With a noninformative prior, each predictor
variable has a probability of $1/p$ of being selected as the splitting
variable for a splitting rule. For the informed prior, a subset of
$p_0$ predictor variables is given twice as much weight, which gives
those variables a larger probability of $2/(p+p_0)$ of being selected
as a splitting variable.

For the fifty data sets generated under each combination of the
parameter settings in the simulations of Section~\ref{secsim-linear},
we implemented three different versions of \texttt{BART}: (1) \texttt
{BART} with a noninformative prior on the predictor variables, (2)
\texttt{BART} with a ``correctly'' informed prior (twice the weight on
the subset of predictor variables that have a true effect on response),
and (3) \texttt{BART} with an ``incorrectly'' informed prior (twice the
weight on a random subset of spurious predictor variables). For each of
these \texttt{BART} models, predictor variables were then selected
using the cross-validated threshold strategy.

\begin{figure}%[H]

\includegraphics{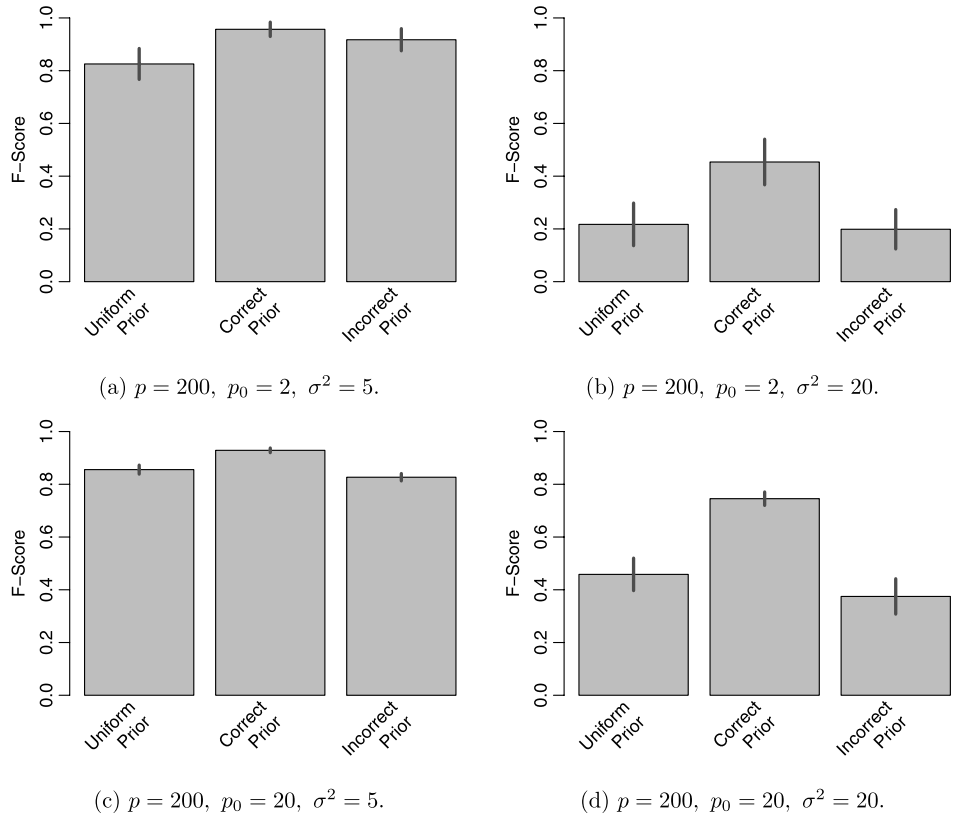}

\caption{Average $F_1$ measures across 50 simulations for \texttt
{BART}-based variable selection under three different prior choices.
The black bars represent 90\% error bars for the average. The settings
shown are the same as those in Figure~\protect\ref{figlin1}\textup{(a)--(d)}.}\label{figF1-prior}\label{figprior1}\label{figprior2}\label{figprior3}\label{figprior4}
\end{figure}

Figure~\ref{figF1-prior} gives the $F_1$ measures for the three
different \texttt{BART} priors in four of the data settings outlined
in Section~\ref{secsim-linear}.

%
%f6 #&#

There are two key observations from the results in Figure~\ref
{figF1-prior}. First, correct prior information can substantially
benefit the variable selection ability of our \texttt{BART}
adaptation, especially in higher noise settings [Figure~\ref{figprior2}(b)~and~(d)]. Second, incorrect prior
information does not degrade performance in any of the cases, which
suggests that our \texttt{BART}-based variable selection procedure is
robust to the misspecification of an informed prior on the predictor
variables. This seems to be a consequence of the Metropolis--Hastings
step, which tends to not accept splitting rules that substantially
reduce the model's posterior value, regardless of how often they are proposed.

To summarize our simulation studies in Section~\ref{secsimulations},
our \texttt{BART}-based variable selection procedure is competitive
with alternative approaches when there is a linear relationship between
the predictor variables and the response, and performs better than
alternative approaches in a nonlinear data setting. \texttt
{BART}-based variable selection can be further improved by correctly
specifying prior information (when available) that gives preference to
particular predictor variables and appears to be robust to
misspecification of this prior information.

%s5 #&#
\section{Application to gene regulation in yeast}\label{secgeneapp}

Experimental advances in molecular biology have led to the availability
of high-dimensional genomic data in a variety of biological
applications. We will apply our \texttt{BART}-based variable selection
methodology to infer the gene regulatory network in budding yeast (\emph{Saccharomyces cerevisiae}). One of the primary mechanisms by which
genes are regulated is through the action of transcription factors,
which are proteins that increase or decrease the expression of a
specific set of genes.

The data for our analyses are expression measures for 6026 genes in
yeast across 314 experiments. For those same 314 experiments, we also
have expression measures for 39 known transcription factors. For each
of the 6026 genes, our goal is to identify the subset of the 39
transcription factors that have a real regulatory relationship with
that particular gene.

We consider each of the 6026 genes as a separate variable selection
problem. For a particular gene $g$, we model the expression measures
for that gene as a $314 \times1$ response vector $\by_g$ and we have
39 predictor variables ($\mathbf{x}_1,\ldots,\mathbf{x}_{39}$) which
are the expression measures of each of the 39 transcription factors.
This same data was previously analyzed using a linear regression
approach in \citet{JenCheSto07}, but we will avoid assumptions of
linearity by employing our \texttt{BART}-based variable selection procedure.

We also have additional data available for this problem that can be
used as prior information on our predictor variables. \citet
{Lee2002} performed chromatin immunoprecipitation (ChIP) experiments
for each of the 39 transcription factors that we are using as predictor
variables. The outcome of these experiments is the estimated
probabilities $m_{gk}$ that gene $g$ is physically bound by each
transcription factor $k$. \citet{CheJenSto07} give details on how
these probabilities $m_{gk}$ are derived from the ChIP data.\footnote
{Probabilities were truncated to be between 5\% and 95\%.}

We will incorporate these estimated probabilities into our \texttt
{BART}-based variable selection approach as prior information. When
selecting predictor variables for splitting rules, we give more weight
to the transcription factors $k$ with larger prior probabilities
$m_{gk}$ in the \texttt{BART} model for gene $g$. Specifically, we
have a splitting variable weight $w_{gk}$ for predictor $\mathbf{x}_k$
in the \texttt{BART} model for gene $g$, which we calculate as
%
%
%e9 #&#
\begin{equation}
\label{eqpriorweights} w_{gk}=1+ c\cdot m_{gk}.
\end{equation}

In the \texttt{BART} model for gene $g$, each predictor $\mathbf
{x}_k$ is chosen for a splitting rule with probability proportional to
$w_{gk}$. The global parameter $c$ controls how influential the
informed prior probabilities $m_{gk}$ are on the splitting rules in
\texttt{BART}. Setting $c=0$ reduces our informed prior to the uniform
splitting rules of the standard \texttt{BART} implementation. Larger
values of $c$ increase the weights of predictor variables with large
prior probabilities $m_{gk}$, giving the informed prior extra influence.

In a real data setting such as our yeast application, it is difficult
to know how much influence to give our informed priors on the predictor
variables. We will consider several different values of $c ={}$\{0, 1, 2, 4, 10,000\} and choose the value that results in the smallest
prediction error on a subset of the observed data that is held out from
our \texttt{BART} model estimation. Specifically, recall that we have
314 expression measures for each gene in our data set. For each gene,
we randomly partition these observations into an 80\% training set,
10\% tuning set, and 10\% hold-out set. For each value of $c ={}$\{0, 1, 2, 4, 10,000\}, we fit a \texttt{BART} model on the 80\% training
set and then choose the value of $c$ that gives the smallest prediction
error on the 10\% tuning set. This same 10\% tuning set is also used to
choose the best threshold procedure among the three options outlined in
Section~\ref{subsecpermdist}. We will use the terminology ``\texttt
{BART}-Best'' to refer to the \texttt{BART}-based variable selection
procedure that is validated over the choice of $c$ and the three
thresholding strategies. While we could also cross-validate over the
significance level $\alpha$, we fix $\alpha=0.05$ due to computational
concerns given the large number of data sets to be analyzed.

For each gene, we evaluate our approach by refitting \texttt{BART}
using only the variables selected by our \texttt{BART}-based variable
selection model and evaluate the prediction accuracy on the final 10\%
hold-out set of data for that gene. This same 10\% hold-out set of data
for each gene is also used to evaluate the prediction accuracy of
various alternative variable selection methods. We consider the
alternative methods of stepwise regression, lasso regression, \texttt
{RF}, \texttt{DT}, and \texttt{Spike-slab} in similar fashions to
Section~\ref{secsimulations}. The 10\% tuning set is used to choose
the value of the penalty parameter $\lambda$ for lasso regression as
well as the importance score threshold for \texttt{RF}. For \texttt
{DT}, we use a constant leaf model for variable selection and then
construct a linear leaf model using the selected variables for prediction.

We also consider three simpler approaches that do not select particular
predictor variables: (1) ``\texttt{BART}-Full'' which is the \texttt
{BART} model using all variables, (2)~ordinary least squares regression
(OLS) with all predictor variables included, and (3)~the ``null'' model:
the sample average of the response which does not make use of any
predictors. In the null model, we do include an intercept, so we are
predicting for the hold-out set of expression measures for each gene
with the average expression level of that gene in the training set.

We first examined the distribution of RMSEs across the 6026 genes. We
found that each procedure improves over the null model with no
covariates, suggesting that some subset of transcription factors is
predictive of gene expression for most of the 6026 genes. However, for
a minority of genes, the null model is competitive, suggesting that the
39 available transcription factors may not be biologically relevant to
every one of these genes. The nonnull variable selection methods show
generally similar performance in terms of the distribution of RMSEs,
and a corresponding figure can be found in the supplementary materials
[\citet{Bleich2014}]. It is important to note that predictive accuracy
in the form of out-of-sample RMSE is not the most desirable metric for
comparing variable selection techniques because it overweights recall
relative to precision.

%
%f7 #&#
\begin{figure}[b]

\includegraphics{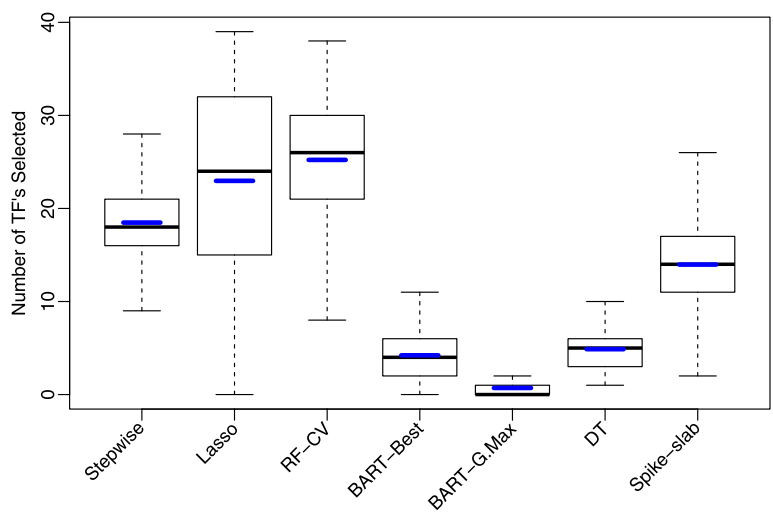}

\caption{Distributions of the number of predictor variables selected
for each method across all 6026 genes. Blue bars represent the average
number of selected predictor variables. Not shown are the null model
which uses no predictors as well as OLS and the full \texttt{BART}
model which both use all predictors. Points beyond the whiskers are omitted.}
\label{fignumvar}
\end{figure}

In Figure~\ref{fignumvar}, we show the distribution of the number of
selected predictor variables (TFs) across the 6026 genes, where we see
substantial differences between the variable selection procedures.
Figure~\ref{fignumvar} confirms that \texttt{BART}-G.max is
selecting very few TFs for each gene. Even more interesting is the
comparison of \texttt{BART}-Best to stepwise regression, lasso
regression, \texttt{RF}, and \texttt{Spike-slab}. \texttt{BART}-Best
is selecting far fewer TFs than these alternative procedures.
Interestingly, \texttt{DT}, the other Bayesian tree-based algorithm,
selects a number of TFs most comparable to \texttt{BART}-Best.

Given the relatively similar performance of methods in terms of RMSE
and the more substantial differences in number of variables selected,
we propose the following combined measure of performance for each
variable selection method:
\[
(\mbox{RMSE reduction per predictor})_\mathrm{method} =
\frac{\mathrm{RMSE}_\mathrm{null} - \mathrm{RMSE}_\mathrm{method}}{\mathrm{NumPred}_\mathrm{method}},
\]
where $\mathrm{RMSE}_\mathrm{method}$ and $\mathrm{NumPred}_\mathrm{method}$ are, respectively, the out-of-sample RMSE and number of
predictors selected for a particular method. This performance metric
answers the question: how much
``gain'' are we getting for adding each predictor variable suggested by
a variable selection approach? Methods that give larger RMSE reduction
per predictor variable are preferred.

%
%f8 #&#
\begin{figure}[b]

\includegraphics{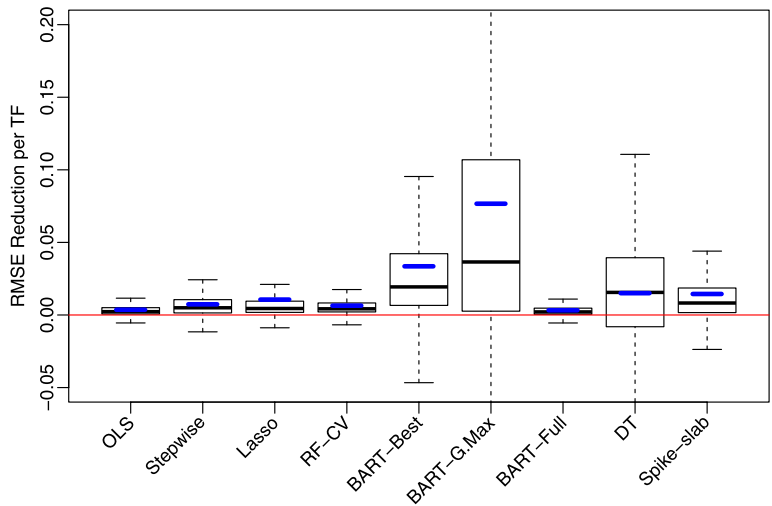}

\caption{Distributions of the RMSE reduction per predictor for each
method across all 6026 genes. Blue bars represent the average RMSE
reduction per predictor. Points beyond the whiskers are omitted.}
\label{figrmsepervar}
\end{figure}

Figure~\ref{figrmsepervar} gives the RMSE reduction per predictor
for each of our variable selection procedures. Note that we only plot
cases where at least one predictor variable is selected, since RMSE
reduction per predictor is only defined if the number of predictors
selected is greater than zero.

Our \texttt{BART}-Best variable selection procedure gives generally
larger (better) values of the RMSE reduction per predictor measure than
stepwise regression, lasso regression, \texttt{RF}, and \texttt
{Spike-slab}. \texttt{DT} is the closer competitor, but does slightly
worse, on average, than \texttt{BART}-Best. Also, both the \texttt
{BART}-Full and OLS procedures, where no variable selection is
performed, perform worse than the variable selection procedures.

\texttt{BART}-G.max, the \texttt{BART}-based procedure under the
global max threshold, seems to perform even better than the \texttt
{BART}-Best procedure in terms of the RMSE reduction per predictor
measure. However, recall that we are plotting only cases where at least
one predictor was selected. \texttt{BART}-G.max selects at least one
transcription factor for only 2866 of the 6026 genes, though it shows
the best RMSE reduction per predictor in these cases. By comparison,
\texttt{BART}-Best selects at least one transcription for 5459 of the
6026 genes while showing better RMSE reduction per predictor than the
non-\texttt{BART} variable selection procedures.

Additionally, Table~\ref{tabpriorvaldist} shows the proportion of
times each choice of prior influence $c$ appeared in the ``\texttt
{BART}-Best'' model. Almost a quarter of the time, the prior
information was not used. However, there is also a large number of
genes for which the prior was considered to have useful information and
was incorporated into the procedure.

%
%t2 #&#
\begin{table}%[htp]
\tabcolsep=0pt
\tablewidth=170pt
\caption{Distribution of prior influence values $c$ used across the 6026 genes}\label{tabpriorvaldist}
\begin{tabular*}{\tablewidth}{@{\extracolsep{\fill}}@{}lc@{}}
\hline
$\bolds{c}$ \textbf{value} & \textbf{Percentage of genes}\\
\hline
\phantom{0000,}0 & 23.3\% \\
\phantom{0000,}0.5 & 16.1\% \\
\phantom{0000,}1 & 15.4\% \\
\phantom{0000,}2 & 14.9\% \\
\phantom{0000,}4 & 14.6\% \\
10,000 & 15.7\% \\
\hline
\end{tabular*}
\end{table}

\citet{JenCheSto07} also used the same gene expression data (and
ChIP-based prior information) to infer gene--TF regulatory
relationships. A~direct model comparison between our \texttt
{BART}-based procedures and their approach is difficult since \citet{JenCheSto07} fit a simultaneous model across all genes, whereas our
current \texttt{BART}-based analysis fits a predictive model for each
gene separately. In both analyses, prior information for each gene--TF
pairing from ChIP binding data [\citet{Lee2002}] was used.\footnote
{\citet{JenCheSto07} used additional prior information based on
promoter sequence data that we did not use in our analysis.}
However, in \citet{JenCheSto07} the prior information for a particular
TF was given the same weight (relative to the likelihood) for each gene
in the data set. In our analysis, each gene was analyzed separately and
so the prior information for a particular TF can be weighted
differently for each gene.

A result of this modeling difference is that the prior information
appears to have been given less weight by our \texttt{BART}-based
procedure across genes, as evidenced by the substantial proportion of
genes in Table~\ref{tabpriorvaldist} that were given zero or low
weight ($c = 0$ or $c=0.5$). Since that prior information played the
role in \citet{JenCheSto07} of promoting sparsity, a consequence of
that prior information being given less weight in our \texttt
{BART}-based analysis is reduced promotion of sparsity.

%
%f9 #&#
\begin{figure}

\includegraphics{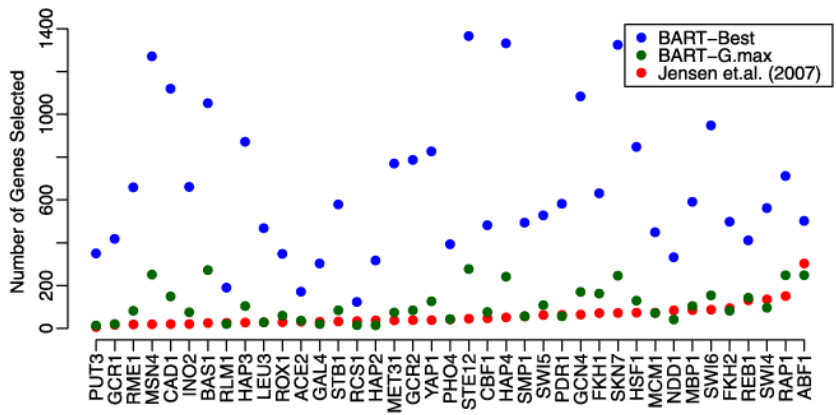}

\caption{Number of genes for which each TF was selected. Results are
compared for \texttt{BART}-Best, \texttt{BART}-G.Max, and the linear
hierarchical model developed in \citet{JenCheSto07}.}
\label{figyeastcomparison}
\end{figure}

This consequence is evident in Figure~\ref{figyeastcomparison},
where we compare the number of selected TFs. The $x$-axis gives the 39
transcription factors that served as the predictor variables for each
of our 6026 genes. The $y$-axis is the number of genes for which that TF
was selected as a predictor variable by each of three procedures:
\texttt{BART}-Best, \texttt{BART}-G.max, and the analysis of \citet{JenCheSto07}. The most striking feature of Figure~\ref
{figyeastcomparison} is that each TF was selected for many more genes
under our \texttt{BART}-Best procedure compared to \texttt
{BART}-G.max, which also selected more variables than the analysis of
\citet{JenCheSto07}. This result indicates that selecting more TFs per
gene leads to the best out-of-sample predictive performance (i.e.,
\texttt{BART}-Best). It could be that \citet{JenCheSto07} were
over-enforcing sparsity, but that previous method also differed from
our current approach in terms of assuming a linear relationship between
the response and predictor variables.

%s6 #&#
\section{Conclusion}\label{secdiscussion}

Chipman, George and
McCulloch's (\citeyear{Chipman10}) Bayesian Additive Regression Trees is a rich
and flexible model for estimating complicated relationships between a
response variable and a potentially large set of predictor variables.
We adapt \texttt{BART} to the task of variable selection by employing
a permutation procedure to establish a null distribution for the
variable inclusion proportion of each predictor. We present several
thresholding strategies that reflect different beliefs about the degree
of sparsity among the predictor variables, as well as a
cross-validation procedure for choosing the best threshold when the
degree of sparsity is not known {a~priori}.

In contrast with popular variable selection methods such as stepwise
regression and lasso regression, our \texttt{BART}-based approach does
not make strong assumptions of linearity in the relationship between
the response and predictors. We also provide a principled means to
incorporate prior information about the relative importance of
different predictor variables into our procedures.

We used several simulated data settings to compare our \texttt
{BART}-based approach to alternative variable selection methods such as
stepwise regression, lasso regression, random forests, and dynamic
trees. Our variable selection procedures are competitive with these
alternatives in the setting where there is a linear relationship
between response and predictors, and performs better than these
alternatives in a nonlinear setting. Additional simulation studies
suggest that our procedures can be further improved by correctly
specifying prior information (if such information is available) and
seem to be robust when the prior information is incorrectly specified.

We applied our variable selection procedure, as well as alternative
methods, to the task of selecting a subset of transcription factors
that are relevant to the expression of individual genes in yeast (\emph{Saccharomyces cerevisiae}). In this application, our \texttt
{BART}-based variable selection procedure generally selected fewer
predictor variables while achieving similar out-of-sample RMSE compared
to the lasso and random forests. We combined these two observations
into a single performance measure, RMSE reduction per predictor. In
this application to inferring regulatory relationships in yeast, our
\texttt{BART}-based variable selection demonstrates much better
predictive performance than alternative methods such as lasso and
random forests while selecting more transcription factors than the
previous approach of \citet{JenCheSto07}.

While we found success using the variable inclusion proportions as the
basis for our procedure, fruitful future work would be to explore the
effect of a variance reduction metric, such as that explored in
\citet{Gramacy2013} within \texttt{BART}.\newpage

%
%sA #&#
\begin{appendix}\label{thresholdalgorithms}
\section*{Appendix: Pseudo-code for variable selection procedures}

\begin{algorithm}%[H]
\caption{Local threshold procedure}\label{alglocalselection}
\begin{algorithmic}
\State Compute $p_1,\ldots,p_K$ \Comment{Inclusion proportions from
original data}
\For{$i \gets\{1,\ldots,P\}$} \Comment{$P$ is the number of null permutations}
\State$\mathbf{y^*}\gets\mathrm{Permute}(\mathbf{y})$
\State Run \texttt{BART} using $\mathbf{y^*}$ as response
\State Compute $p_{i1}^*,\ldots,p_{iK}^*$ \Comment{Inclusion proportions
from permuted data}
\EndFor
\For{$j \gets\{1,\ldots,K\}$}
\State$q^*_j \gets\operatorname{Quantile}(p^*_{1j},\ldots,p^*_{Pj},1-\alpha
)$ \Comment{$1-\alpha$ quantile\vspace*{2pt} of $\mathbf{x}_j$ permutation distribution}
\State\textbf{if} $p_j > q^*_j$ \textbf{then} Include $p_j$ in Vars
\textbf{end if}
\EndFor
\State\Return{Vars}
\end{algorithmic}
\end{algorithm}\vspace*{-6pt}

\mbox{}

\begin{algorithm}%[p]
\caption{Global maximum threshold procedure}\label{algglobalmax}
\begin{algorithmic}
\State Compute $p_1,\ldots,p_K$ \Comment{Inclusion proportions from
original data}
\For{$i \gets\{1,\ldots,P\}$} \Comment{$P$ is the number of null permutations}
\State$\mathbf{y^*}\gets\mathrm{Permute}(\mathbf{y})$
\State Run \texttt{BART} using $\mathbf{y^*}$ as response
\State Compute $p_{i1}^*,\ldots,p_{iK}^*$ \Comment{Inclusion proportions
from permuted data}
\State$g_i \gets\mathrm{Max}(p_{i1}^*,\ldots,p_{iK}^*)$ \Comment
{Maximum of proportions from permuted data}
\EndFor
\State$g^* \gets\operatorname{Quantile}(g_i,\ldots,g_P,1-\alpha)$ \Comment
{$1-\alpha$ Quantile of maxima}
\For{$j \gets\{1,\ldots,K\}$}
\State\textbf{if} $p_j > g^*$ \textbf{then} Include $p_j$ in Vars
\textbf{end if}
\EndFor
\State\Return{Vars}
\end{algorithmic}
\end{algorithm}
\vspace*{20pt}

\begin{algorithm}%[p]
\caption{Global standard error threshold procedure}\label{algglobalse}
\begin{algorithmic}
\State Compute $p_1,\ldots,p_K$ \Comment{Inclusion proportions from
original data}
\For{$i \gets\{1,\ldots,P\}$} \Comment{$P$ is the number of null permutations}
\State$\mathbf{y^*}\gets\mathrm{Permute}(\mathbf{y})$
\State Run \texttt{BART} using $\mathbf{y^*}$ as response
\State Compute $p_{i1}^*,\ldots,p_{iK}^*$ \Comment{Inclusion proportions
from permuted data}
\EndFor
\For{$j \gets\{1,\ldots,K\}$}
\State$m_j \gets\mathrm{Avg}(p^*_{1j},\ldots,p^*_{Pj})$ \Comment
{Sample average of $\mathbf{x}_j$ permutation distribution}
\State$s_j \gets\mathrm{SD}(p^*_{1j},\ldots,p^*_{Pj})$ \Comment{Sample
sd of $\mathbf{x}_j$ permutation distribution}
\EndFor
\State$C^* \gets\inf_{C\in\mathbb{R}^+} \{
\forall j$, $\frac{1}{P}\sum_{i=1}^P\mathbb{I} (p^*_{ij} \leq
m_j + C \cdot s_j ) > 1 - \alpha\}$ \Comment
{Simultaneous coverage} %\displaystyle\min_{j \in\bracesss{1,\ldots,K}}
\For{$j \gets\{1,\ldots,K\}$}
\State\textbf{if} $p_j > m_j + C^*\cdot s_j$ \textbf{then} Include
$p_j$ in Vars \textbf{end if}
\EndFor
\State\Return{Vars}
\end{algorithmic}
\end{algorithm}\vspace*{25pt}

\begin{algorithm}%[H]
\caption{Cross-Validated Comparison of Threshold Procedures}\label{algcvthree}
\begin{algorithmic}
\State Divide the data into $K$ training-test splits
\For{$k \gets\{1,\ldots,K\}$}
\For{method $ \gets\{$Local, Global Maximum, Global SE$\}$}
\State$\operatorname{Var}_{\mathrm{method}} \gets$ Selected variables
using method on \texttt{BART}
\State$\mathrm{\texttt{BART}}_{\mathrm{method}} \gets$ \texttt
{BART} built from $k$th training set using only $\operatorname{Var}_{\mathrm{method}}$
\State$L2_{k, \mathrm{method}} \gets$ $L2$ error from
$\mathrm{\texttt{BART}}_{\mathrm{method}}$ on $k$th test set
\EndFor
\EndFor
\For{method $ \gets\{$Local, Global Maximum, Global SE$\}$}
\State$L2_{\mathrm{method}} \gets\sum_{k=1}^K L2_{k,
{\mathrm{method}}}$ \Comment{Aggregate $L2$ error over entire
training set}
\EndFor
\State method$^* \gets\arg\min_{\mathrm{method}}
\{L2_{\mathrm{method}} \}$ \Comment{Choose the best method from
the three}
\State\Return Selected variables using method$^*$ for \texttt{BART}
on full training set
\end{algorithmic}
\end{algorithm}
\end{appendix}

\section*{Acknowledgments}
%Adam Kapelner acknowledges the National Science Foundation for his graduate research fellowship.
We would like to thank Hugh MacMullan for
help with grid computing and the anonymous reviewers for their
insightful comments.

\begin{supplement}[id=suppA]
\stitle{Additional results for simulations and gene regulation application}
\slink[doi]{10.1214/14-AOAS755SUPP} %[doi,text={...}] - jei reikia suskaldyti doi
\sdatatype{.pdf}
\sfilename{aoas755\_supp.pdf}
\sdescription{Complete set of results for simulations in Section~\ref{secsimulations} and additional output for Section~\ref{secgeneapp}.}
\end{supplement}

% imsref loaded by linak, 2014-07-30 13:51:13
%
% imsref loaded by linak, 2014-08-05 14:26:22

\printaddresses
\end{document}